\documentclass[
prd,amsmath]{revtex4-1}

\renewcommand{\Im}{\textrm{Im}\,}
\newcommand{\tr}{\textrm{tr}\,}
\newcommand{\CP}{\textit{CP}}
\newcommand{\abs}[1]{\left| #1 \right|}
\newcommand{\dm}[2]{(m^2_{#1}-m^2_{#2})}
\newcommand{\dmu}[2]{(m^2_{#1}-m^2_{#2})}

\newcommand{\Vsq}[1]{\left| V_{#1} \right|^2}
\newcommand{\V}[3]{V_{{#1}{#2}}V^{*}_{{#1}{#3}}}
\newcommand{\NTU}{Department of Physics, National Taiwan University,
    Taipei, Taiwan 10617}
\newcommand{\NCTSn}{National Center for Theoretical Sciences, North Branch,
    National Taiwan University, Taipei, Taiwan 10617}

\begin{document}

\title{Leading Effect of \CP~Violation with Four Generations}

\author{Wei-Shu Hou$^{1,2}$, Yao-Yuan Mao$^{1}$, and Chia-Hsien Shen$^{1}$\\
{$^{1}$ \textit \NTU}\\
{$^{2}$ \textit \NCTSn}
 }

\begin{abstract}
In the Standard Model with a fourth generation of quarks, we study
the relation between the Jarlskog invariants and the triangle
areas in the $4\times 4$ CKM matrix. To identify the leading
effects that may probe the \CP~violation in processes involving
quarks, we invoke small mass and small angle expansions, and show
that these leading effects are enhanced considerably compared to
the three generation case by the large masses of fourth generation
quarks. We discuss the leading effect in several cases, in
particular the possibility of large \CP~violation in $b\to s$
processes, which echoes the heightened recent interest because of
experimental hints.
\end{abstract}

\maketitle

\section{Introduction and Motivation}

The experimental discovery~\cite{CPV1964} of \CP~Violation (CPV)
in 1964 came as a surprise, but it provided the clue that lead
Sakharov to suggest~\cite{Sakharov} the three conditions that need
to be satisfied to explain a long standing and profound puzzle:
the disappearance of antimatter from the very early Universe. It
was Kobayashi and Maskawa (KM) who proposed~\cite{KM}, in 1973,
that CPV can arise from charged current weak interactions, if
there is a third generation of quarks. This proposal became part
of the Standard Model (SM). The $3\times 3$ quark mixing matrix
can describe all flavor physics measurements to date, the crowning
glory being the measurement of the fundamental CPV phase in modes
such as $B^0\to J/\psi K_S$ by the B factory
experiments~\cite{PDG}. However, even the Nobel committee
noted~\cite{Nobel} that the KM phase is insufficient for the
Sakharov conditions, typically by a factor of $10^{-10}$ or worse.

The numerics for the  $10^{-10}$ factor can be most easily seen by
a dimensional analysis of the so-called Jarlskog invariant CPV
measure~\cite{Jarlskog85, Jarlskog87} of the 3-generation Standard
Model (SM3),
\begin{equation}
J = \dm{t}{u} \dm{t}{c} \dm{c}{u}
    \dm{b}{d} \dm{b}{s} \dm{s}{d} \, A,
 \label{eq:JinSM3}
\end{equation}
and comparing with the baryon-to-photon ratio $n_B/n_\gamma$ of
our Universe, which is of order $10^{-9}$. In
Eq.~(\ref{eq:JinSM3}), $A \simeq 3\times 10^{-5}$~\cite{PDG} is
the area of, e.g. the triangle formed by the three sides of the
unitarity relation $V_{ud}^*V_{ub} + V_{cd}^*V_{cb} +
V_{td}^*V_{tb} = 0$ of SM3. Note that all possible triangle areas
have the same value in SM3. Since $J$ has 12 mass dimensions,
normalizing by say $v$, the vacuum expectation value of
electroweak symmetry breaking (EWSB), or more trivially the
electroweak phase transition temperature $T_{\rm EW} \simeq 100$
GeV, one finds $J/T_{\rm EW}^{12} \lesssim 10^{-20}$ falls short
of $n_B/n_\gamma$ by at least $10^{-10}$.

The main suppression factor of $J$ come mainly from the small
masses, $m_s^2m_c^2m_b^4(/T_{\rm EW}^8)$, rather than from $A$.
Noting this, one of us had suggested~\cite{Hou09} that, \emph{if
one had an extra generation of quarks}, i.e. 4-generation Standard
Model (SM4), then an analog to $J$ in SM3,
\begin{equation}
J_{(2,3,4)}^{sb} = \dm{t'}{c} \dm{t'}{t} \dm{t}{c}
           \dm{b'}{s} \dm{b'}{b} \dm{b}{s} A_{234}^{sb},
 \label{eq:J234_Hou}
\end{equation}
would be enhanced by $\sim$ 15 orders of magnitude with respect to
$J$. The staggering enhancement is brought about by the heavy
$t'$, $b'$ quark masses, which are taken to be in the range of 300
to 600 GeV. A phenomenological analysis~\cite{HNS05} for
$A_{234}^{sb} = \Im \left[ (\V{t}{s}{b})^* \V{t'}{s}{b}\right] $
has been taken into account. It was further argued~\cite{Hou09}
that the proximity to the $m_d \cong m_s \cong 0$ degeneracy
limit~\cite{Jarlskog87} on the scale $v$ implies an almost
3-generation world involving 2-3-4 generation quarks, in which
Eq.~(\ref{eq:J234_Hou}) is indeed the leading term.
Unlike the case for SM3, this large enhancement likely allows SM4
to provide sufficient CPV for the matter or baryon asymmetry of
the Universe (BAU), which provides strong support for the possible
existence of a 4th generation. The scenario can be directly
searched for quite definitely~\cite{Arhrib06} at the LHC.
In fact, because of recent experimental activities at the
Tevatron, be it direct search for the $t'$~\cite{CDFtp} or
$b'$~\cite{CDFbp} quarks, or CPV studies in $B_s \to
J/\psi\,\phi$~\cite{psiphi, psiphiCDF10} or recent hint of dimuon
asymmetry~\cite{ASLDzero}, interest in the 4th generation has been
steadily growing~\cite{HHHMSU,recent}. Our interest here, however,
is the more fundamental.

It may still be questioned whether the analogy of
Eq.~(\ref{eq:J234_Hou}) and Eq.~(\ref{eq:JinSM3}) between SM4 and
SM3 covers the whole truth. Indeed, the expression of invariants
probing the \CP~nonconservation become much more complicated when
one adds a fourth generation, and one should inspect all the
invariant quantities in SM4 more carefully. In the following
section, we first review the various discussions of necessary and
sufficient conditions for CPV. We then invoke the ``natural
ordering" --- the apparent hierarchy of mass-mixing parameters in
the quark sector --- to make a small mass expansion. The fact that
mixing angles involving the 4th generation cannot be large is less
useful, since the pattern of mixing angles (e.g. $\vert
V_{t's}\vert$ vs $\vert V_{t'b}\vert$) is less clear at the
moment, precisely because of the recent hints for possibly large
CPV effects in $b\to s$ transitions. However, we are able to
identify, from the phenomenological indication that $b\to d$
transitions appear consistent with SM3, the condition that
simplifies the Jarlskog invariants further, and confirm that, in
our world, Eq.~(\ref{eq:J234_Hou}) is indeed (close to) the
leading effect for CPV. In the above process, we are also able to
find the next-to-leading terms. We offer some discussion on the
approximations made, before giving our conclusion. More tedious
algebra and a discussion on the relations between triangle areas
are given as appendices.

\section{
         Conditions for \CP~conservation}

Many studies has been made on the necessary and sufficient
conditions for \CP~conservation with 3 and 4 (or more)
generations. In SM3, which is a very special case, we have only
one condition for \CP~conservation:
\begin{equation}
\frac{1}{6}\, \Im \tr [S,\, S']^3 =  -\Im \tr (S^2S'SS'^2) =
J(1,2,3) =  v(1,2,3) v'(1,2,3) A = 0,
 \label{eq:conditionSM3}
\end{equation}
where $ S^{(\prime)} $ is the up(down)-type Hermitian squared mass
matrix, defined as $S = MM^\dagger$ with $M$ the quark mass
matrix. All the primed symbols hereafter denote down-type
quantities. In Eq.~(\ref{eq:conditionSM3}), $v$ is the Vandermonde
determinant of squared masses,
\begin{equation}
v(\alpha,\beta,\gamma) =
    \dmu{\alpha}{\beta} \dmu{\beta}{\gamma}\dmu{\gamma}{\alpha},
 \label{eq:v_def}
\end{equation}
and $J(1,2,3)$ (which is identical to $J$ in Eq.~(\ref{eq:JinSM3})) is
the Jarlskog invariant,
\begin{equation}
J(\alpha,\beta,\gamma) = v(\alpha,\beta,\gamma) \, \Im\tr(P_\alpha
S' P_\beta S' P_\gamma S'),
 \label{eq:J_def}
\end{equation}
where $P_\alpha$ is the projection operator for the indicated
flavor, $SP_\alpha = m_\alpha^2 P_\alpha$.

That the number of conditions for \CP~conservation in SM3 is
exactly one reflects the unique CPV phase in the quark mixing
matrix. With fourth generations, we have two more CKM phases,
hence more conditions are needed for \CP~conservation. The number
of conditions may, however, be larger than 3 due to the complexity
of, and interdependency between, invariants. For instance, Botella
and Chau~\cite{Botella86} showed that there are nine independent
triangle areas in SM4, rather than the single area in SM3, and
\CP~is conserved \emph{if and only if} all nine areas vanish. Note
that these triangles are not ``unitary triangles,'' since in SM4
the unitarity relations give quadrangles. However, every two sides
of these quadrangles still form triangles, and we refer to these
triangles as ``CKM triangles.'' Namely,
\begin{equation}
{A^{u_1u_2}_{d_1d_2}} \equiv
    \Im \left[ (\V{u_1}{d_1}{d_2})^* \V{u_2}{d_1}{d_2}\right] = 0, \quad
    \forall \, u_1 \neq u_2, d_1 \neq d_2,
 \label{eq:A_def}
\end{equation}
is defined quite in the same way with the conventions in SM3. One
can see the number of total possible triangles is $(C^4_2)^2 = (4!/2!2!)^2 =
36 $, but the unitarity conditions reduce this number to
$(C^3_2)^2 = (3!/2!)^2 = 9$ (the corresponding numbers for SM3 are therefore
9 and 1, respectively). These CKM triangles, though rephasing invariant,
may not be fully independent of each other (see Appendix B for some discussion).
Furthermore, quark masses do not appear explicitly, although we know that
CPV would vanish under certain mass degeneracy conditions.

Eq.~(\ref{eq:conditionSM3}), which gives the Jarlskog invariant
CPV measure for SM3 as in Eq.~(\ref{eq:JinSM3}), is of course
invariant under any change of flavor basis. Extending to SM4,
Jarlskog showed~\cite{Jarlskog87} that it is the sum over four Jarlskog invariants
of the form in Eq.~(\ref{eq:J_def}), or three-cycles, that is
\begin{equation}
- \Im \tr (S^2 S' S S'^2) =
     J(2,3,4) + J(1,3,4) + J(1,2,4) + J(1,2,3).
 \label{eq:Jarlskog_condition}
\end{equation}
The Jarlskog proposal is that one would have \CP~conservation
\emph{if and only if} all four invariants vanish.
This proposal shows a transparent analogy between SM3 and SM4.

There are other basis-independent approaches, however, to the
conditions for \CP~conservation. Gronau, Kfir, and Loewy (GKL)
introduced~\cite{Gronau86} 5 more invariants in addition to
Eq.~(\ref{eq:conditionSM3}), and proposed that \CP~is conserved in
SM4 \emph{if and only if} all six invariants vanish,
\begin{equation}
\begin{split}
 & \Im \tr (S^2 S' S S'^2) = \Im \tr(S^2 S' S S'^3) = \Im \tr( S^2 S'^2 S S'^3) \\
=\; & \Im\tr(S' S S'^2 S S'^3) = \Im\tr(S^3 S' S S'^2) =
    \Im\tr(S^3 S' S S'^3) = 0.
\end{split}
 \label{eq:Gronau_condition}
\end{equation}
Whether these two sets of conditions are really 
sufficient for \CP~conservation has been
debated~\cite{Jarlskog87,Gronau89,Jarlskog89}.
What is certain is that, if \CP~is violated, some of these
quantities would be nonzero.
Both sides do agree that {\it the two sets of conditions are
equivalent if there is no vanishing element in the quark mixing
matrix $V$ in SM4}.

The pragmatic question is how these quantities appear in process
amplitudes that give rise to \emph{observable} measures of
\CP~violation. In SM3 we know that the Jarlskog invariant enters
various CPV measures. It further encodes the notion that, if any
two like-charge quarks are degenerate in mass, or if the CKM
triangle area vanishes, there would be no \CP~violation. Taking
this as a hint, it is clear that the CKM triangle areas should
enter various CPV measures, together with some mass difference
factors, so the GKL and Jarlskog invariants should play a role in
these measures. The generation labels of the CKM triangles in the
measure should tell us what are the related processes for the
search of \CP~violation. Note that the GKL and Jarlskog invariants
are basis-independent, and thus more likely to appear in physical
quantities.

The invariants in Eqs.~(\ref{eq:J_def}) and
(\ref{eq:Gronau_condition}) are, however, rather complicated, and
it is not apparent how the fourth generation effect on
\CP~violation emerges. In contrast, the suggested leading effect
of Eq.~(\ref{eq:J234_Hou}) is much more intuitive, and rather
similar to the SM3 result of Eq.~(\ref{eq:JinSM3}), but with the
emphasis placed clearly on $b\to s$ transitions. We should try to
express the invariants, whether the GKL~\cite{Gronau86} or
Jarlskog~\cite{Jarlskog87} kind, in terms of mass difference
factors and CKM triangle areas, just like in SM3, and then
identify the leading terms by considering the physical limits,
such as physical quark masses. An extensive work that expands the GKL
invariants into the nine CKM triangles~\cite{Botella86} has been
done in Ref.~\cite{delAguila96}, but no comparison between
different terms were made. For example, one has,
\begin{equation}
\Im \tr (S^2 S' S S'^2) = \sum v(1,i,j) v'(1,a,b) A^{ij}_{ab},
\end{equation}
where the summation is over $(i,j)$ and $(a,b) \in
\{(2,3),(3,4),(4,2)\}$, with similar expansions for the other
invariants in Eq.~(\ref{eq:Gronau_condition}). Comparing different
GKL invariants would be less meaningful, however, as there are
several different mass dimensions. Even if they are shown to enter
some CPV measure, it would be difficult to tell whether the fourth
generation enhances the \CP~violation or not. On the other hand,
all the Jarlskog invariants have the same dimension, and the
general form of Eq.~(\ref{eq:J_def}) is maintained as one extends
from SM3 to SM4.

In the following, we will express the Jarlskog invariants in terms
of the nine CKM triangles, then identify the leading terms in
these invariants, by adopting proper physical limits. We find that
the suggestion of Eq.~(\ref{eq:J234_Hou}) is indeed the leading
term, but various next-to-leading terms are only
smaller by roughly a factor of 10.

\section{ Small Mass and Angle Expansion of Jarlskog Invariants }

\subsection{ Jarlskog Invariants and CKM triangles }

Let us express the Jarlskog invariants of Eq.~(\ref{eq:J_def})
in terms of the nine CKM triangles of Eq.~(\ref{eq:A_def}).
For convenience, we choose to decompose the down-type Jarlskog
invariants into CKM triangle areas. The up-type relations can be
obtained analogously. Since the invariants can be evaluated in any
basis, we are allowed to choose $S'$ to be diagonal and write
$\hat{S} = V^\dagger M V$, where $V$ is the familiar quark mixing
matrix. The Jarlskog invariants become
\begin{equation}
\begin{split}
J'(a,b,c) &= - v'(a,b,c) \, \Im\tr(P'_a S P'_b S P'_c S)
= - v'(a,b,c) \, \Im(\hat{S}_{ab}\hat{S}_{bc}\hat{S}_{ca})\\
&= \,\;\ v'(a,b,c) \sum_{i,j,k =1}^4 m^2_i m^2_j m^2_k \,
              \Im(\V{i}{a}{b}\V{j}{b}{c}\V{k}{c}{a}),
\end{split}
 \label{eq:Jprime_def}
\end{equation}
where the indices $a,b,c$ are chosen all differently within
$\{1,2,3,4\}$, but not summed. With an eye towards Eq.~(\ref{eq:J234_Hou}),
we use the unitarity relation
\begin{equation}
\V{1}{a}{b}=-\V{2}{a}{b}-\V{3}{a}{b}-\V{4}{a}{b} + \delta_{ab},
  \label{eq:eliminate1}
\end{equation}
to eliminate all 1's for the indices that are summed over. We
factor out any real factor that appears in $\Im(\cdots)$ of
Eq.~(\ref{eq:Jprime_def}), and the remaining four CKM matrix
elements then form a CKM triangle area.

After some algebra, which is rendered to
Appendix~\ref{deriveJinA}, the four down-type Jarlskog invariants
are reduced to
\begin{equation}
\begin{split}
J'(2,3,4) & = - v'(2,3,4) \Lambda_{234}, \\
J'(1,3,4) & =\; v'(1,3,4) \left[\Lambda_{234} - v(1,3,4)A^{34}_{34}
                         - v(1,2,3)A^{23}_{34} - v(1,4,2)A^{42}_{34} \right], \\
J'(1,2,3) & =\; v'(1,2,3) \left[\Lambda_{234} - v(1,3,4)A^{34}_{23}
                         - v(1,2,3)A^{23}_{23} - v(1,4,2)A^{42}_{23} \right], \\
J'(1,4,2) & =\; v'(1,4,2) \left[\Lambda_{234} -v(1,3,4)A^{34}_{42}
                         - v(1,2,3)A^{23}_{42} - v(1,4,2)A^{42}_{42}
\right],
\end{split}
 \label{eq:4invariants}
\end{equation}
where
\begin{equation}
\begin{split}
\Lambda_{234} =
\quad \dmu{2}{1}\dmu{3}{1}
\Big[ \quad &\dmu{4}{2}
      \left(\Vsq{22}A^{23}_{34}+\Vsq{23}A^{23}_{42}+\Vsq{24}A^{23}_{23}\right)\\
      -&\dmu{4}{3}
      \left(\Vsq{32}A^{23}_{34}+\Vsq{33}A^{23}_{42}+\Vsq{34}A^{23}_{23}\right)
\Big]\\
+ \; \dmu{3}{1}\dmu{4}{1}
\Big[ \quad &\dmu{2}{3}
      \left(\Vsq{32}A^{34}_{34}+\Vsq{33}A^{34}_{42}+\Vsq{34}A^{34}_{23}\right)\\
      -&\dmu{2}{4}
      \left(\Vsq{42}A^{34}_{34}+\Vsq{43}A^{34}_{42}+\Vsq{44}A^{34}_{23}\right)
\Big]\\
+ \; \dmu{4}{1}\dmu{2}{1}
\Big[ \quad &\dmu{3}{4}
      \left(\Vsq{42}A^{42}_{34}+\Vsq{43}A^{42}_{42}+\Vsq{44}A^{42}_{23}\right)\\
      -&\dmu{3}{2}
      \left(\Vsq{22}A^{42}_{34}+\Vsq{23}A^{42}_{42}+\Vsq{24}A^{42}_{23}\right)
\Big].\\
\end{split}
 \label{eq:Lambda234_def}
\end{equation}
This rather compact form, though still quite complicated, should
be compared with the many terms of Ref.~\cite{delAguila96},
obtained by expanding the 6 GKL invariants~\cite{Gronau86} of
Eq.~(\ref{eq:Gronau_condition}) in terms of the 9 CKM triangle
areas~\cite{Botella86} of Eq.~(\ref{eq:A_def}). We note again that
the GKL invariants are of several different mass dimensions, and
it is not easy to compare the relative importance of the numerous
possible terms. In contrast, all four Jarlskog invariants have the
same mass dimension, hence are more readily compared with one
another. As remarked already, for SM4 with no vanishing CKM mixing
matrix elements, GKL and Jarslkog approaches are equivalent, but
the latter is clearly more convenient for our purpose.

\subsection{ Small Mass and Angle Expansions }

One of our goals is to identify the leading effects in the
Jarlskog invariants. We now depart from generality by noting the
fact of a clear hierarchy of physical quark masses in Nature,
namely $m^2_{b'} \gg m^2_b \gg m^2_s \gg m^2_d $. This implies
that the last two factors in Eq.~(\ref{eq:4invariants}) are much
more suppressed than the first two.
Likewise, the up-type hierarchy $m^2_{t'} > m^2_t \gg m^2_c \gg
m^2_u $ further suppresses the first and third terms in
$\Lambda_{234}$, as well as the terms with factors $v(1,2,3)$ and
$v(1,2,4)$ in $J'(1,3,4)$.
Dropping these $m_s^2$ and $m_c^2$ suppressed terms,
Eqs.~(\ref{eq:4invariants}) and (\ref{eq:Lambda234_def}) become
\begin{equation}
\begin{split}
J'(2,3,4) & \simeq -  \dm{b'}{s} \dm{b'}{b} \dm{b}{s} \Lambda_{234}, \\
J'(1,3,4) & \simeq \; \,\ \dm{b'}{d} \dm{b'}{b} \dm{b}{d}
    \left[\Lambda_{234} - \dm{t'}{u} \dm{t'}{t} \dm{t}{u} A^{tt'}_{bb'}\right],
\end{split}
  \label{eq:2invariants}
\end{equation}
\begin{equation}
\begin{split}
\Lambda_{234} \simeq \; \dm{t}{u}\dm{t'}{u}
\Big[ \quad &\dm{t}{c}
      \left(-\Vsq{ts}A^{tt'}_{bb'}+\Vsq{tb}A^{tt'}_{sb'}-\Vsq{tb'}A^{tt'}_{sb}\right)\\
      +&\dmu{t'}{c}
      \left(\Vsq{t's}A^{tt'}_{bb'}-\Vsq{t'b}A^{tt'}_{sb'}+\Vsq{t'b'}A^{tt'}_{sb}\right)
\Big].
\end{split}
  \label{eq:Lambda234_nomc}
\end{equation}
where we have returned all indices to physical labels, and kept
the small subtracted masses in the explicit mass differences in
these remainder terms, for sake of correspondence with
Eq.~(\ref{eq:J234_Hou}).

As there are still quite a few terms in
Eq.~(\ref{eq:Lambda234_nomc}), we expand further in the strength
of $\Vsq{ij}$. Phenomenologically, we now know~\cite{recent} that
the rotation angles in the CKM matrix are small, i.e. $ \Vsq{ij}
\ll \Vsq{kk} \sim 1 $, which holds not only in SM3, but seems to
extend into SM4 as well. The Cabibbo angle appears to be the
largest rotation angle, while $\vert V_{ts}\vert$ cannot be much
different from the SM3 value of $\simeq 0.04$, and $\Vsq{t's}$ and
$\Vsq{t'b}$ should be of order $10^{-2}$ or less~\cite{Arhrib09}.
Assuming small rotation angles, we drop the off-diagonal $\vert
V_{ij}\vert^2$ terms in $\Lambda_{234}$, and get
\begin{align}
\begin{split}
J'(2,3,4) \sim - &\dm{t'}{u} \dm{t}{u} \dm{b'}{s} \dm{b'}{b} \dm{b}{s} \\
    & \times \left[
     \dm{t}{c} \Vsq{tb}A^{tt'}_{sb'} + \dm{t'}{c} \Vsq{t'b'}A^{tt'}_{sb}
    \right],
  \label{eq:leadinginJ234}
\end{split} \\
\begin{split}
J'(1,3,4) \sim \;\,\ &\dm{t'}{u} \dm{t}{u} \dm{b'}{d} \dm{b'}{b} \dm{b}{d} \\
    & \times \left[
    \dm{t}{c} \Vsq{tb}A^{tt'}_{sb'} +  \dm{t'}{c} \Vsq{t'b'}A^{tt'}_{sb}
   - \dm{t'}{t}  A^{tt'}_{bb'} \right].
  \label{eq:leadinginJ134}
\end{split}
\end{align}
If one looks at the order of magnitude of quark masses,
Eq.~(\ref{eq:leadinginJ234}) is similar to
Eq.~(\ref{eq:J234_Hou}). But they are not exactly the same: more
than one CKM area carry the heaviest mass factor $m_{t'}^4 m_t^2
m_{b'}^4 m_b^2$, and the masses of $u$, $d$ quarks also enter the
expression.
%

In Eqs.~(\ref{eq:leadinginJ234}) and (\ref{eq:leadinginJ134}), we
have used ``$\sim$" rather than ``$\simeq$", because we have
treated the CKM triangle areas as ``free parameters" while
dropping the off-diagonal $\vert V_{ij}\vert^2$ terms. These areas
are characterized not only by the strength of CKM matrix elements,
but also their relative phases, which makes clear that these two
equations are for illustration purpose only. Note that we have not
assumed further hierarchical structure in the mixing elements.
This is because of the possible indication of large CPV effect
involving $b\to s$ transitions, hence we do not know whether the
hierarchy structure of $\vert V_{ub}\vert^2 \ll \vert
V_{cb}\vert^2 \ll \vert V_{us}\vert^2
 \ll 1$ would extend to elements involving the 4th generation.
We remark that, without assuming further structure in the CKM
elements, unlike the application of mass hierarchies that lead to
Eqs.~(\ref{eq:2invariants}) and (\ref{eq:Lambda234_nomc}), had we
applied $ \Vsq{ij} \ll \Vsq{kk} \sim 1 $ (where $i \neq j$) first
to Eqs.~(\ref{eq:4invariants}) and (\ref{eq:Lambda234_def}), not
much simplification would be gained, and there would still be four
Jarlskog invariants.

What CKM angle pattern could be noteworthy for further
simplifications?

\subsection{ Leading Effect of Jarlskog Invariants }

For our purpose of finding the leading effect of CPV, what pattern
of small off-diagonal elements in the CKM matrix could provide
additional approximate relations between triangle areas as defined
in Eq.~(\ref{eq:A_def})? Taking note of the three specific
triangle areas that enter Eq.~(\ref{eq:Lambda234_nomc}), we note
that the unitarity quadrangle
\begin{equation}
V_{td}V^*_{t'd} + V_{ts}V^*_{t's} + V_{tb}V^*_{t'b} + V_{tb'}V^*_{t'b'} = 0,
  \label{eq:unitarity_v41v31}
\end{equation}
a special case of our starting Eq.~(\ref{eq:eliminate1}), would
approach a triangle, if $ \abs{V_{td}V^*_{t'd}} $ is small
compared to the other terms. If this is the case, then any two of
the other three sides would form the same triangle area, i.e.
\begin{equation}
A^{tt'}_{sb} \simeq - A^{tt'}_{sb'} \simeq A^{tt'}_{bb'}, \qquad
{\rm for}\ \abs{V_{td}V^*_{t'd}} \ll 1,
  \label{eq:areas_v41v31}
\end{equation}
which relates the three CKM triangles appearing in $\Lambda_{234}$
of Eq.~(\ref{eq:Lambda234_nomc}).
Applying Eq.~(\ref{eq:areas_v41v31}) to
Eq.~(\ref{eq:Lambda234_nomc}), we get
\begin{equation}
\begin{split}
\Lambda_{234} &\simeq \; \dm{t}{u}\dm{t'}{u}
\Big[ -\dm{t}{c}
      \left(1 - \Vsq{td} \right) A^{tt'}_{sb}
      +\dmu{t'}{c}
      \left(1 - \Vsq{t'd} \right) A^{tt'}_{sb} \Big] \\
&\simeq \; \dm{t}{u}\dm{t'}{u}\dm{t'}{t} A^{tt'}_{sb},
\end{split}
  \label{eq:Lambda234_v41v31}
\end{equation}
where we have assumed the smallness of both $\Vsq{t'd} $ and
$\Vsq{td}$ in the second step, which is somewhat stronger than
what is needed for Eq.~(\ref{eq:areas_v41v31}) to hold.
Substituting Eq.~(\ref{eq:Lambda234_v41v31}) into
Eq.~(\ref{eq:2invariants}), we obtain
\begin{align}
J'(2,3,4) & \simeq  - \dm{b'}{s} \dm{b'}{b} \dm{b}{s} \dm{t}{u}\dm{t'}{u}\dm{t'}{t} A^{tt'}_{sb},
  \label{eq:J234_v41v31} \\
J'(1,3,4) & \simeq    0 + \text{subleading terms}.
  \label{eq:J134_v41v31}
\end{align}

With $A_{234}^{sb} = A^{tt'}_{sb}$, $J'(2,3,4)$ in
Eq.~(\ref{eq:J234_v41v31}) is indeed the same as
$J_{(2,3,4)}^{sb}$ of Eq.~(\ref{eq:J234_Hou}), except that $m_c^2$
is replaced by $m_u^2$, which makes little difference since
$m_u^2$ and $m_c^2$ are negligible compared with $m_t^2$ and
$m_{t'}^2$.
What seems a little curious is that the starting point of
Eq.~(\ref{eq:unitarity_v41v31}) is more relevant for $t' \to t$
transitions, but it leads to the result in
Eq.~(\ref{eq:J234_v41v31}) (through Eq.~(\ref{eq:areas_v41v31})),
which seems more relevant to $b\to s$ transitions.

In Ref.~\cite{Arhrib09}, the authors also showed that $
A^{tt'}_{sb'} $ and $ A^{tt'}_{sb} $ have similar area through a
phenomenological study. Such a study started~\cite{HNS05} from the
hint of New Physics CPV in $b\to s$ transitions, while $b\to d$
transitions (including $B_d$ mixing-dependent CPV) mimic SM3, as
first noted by Ref.~\cite{HNSprd05}. If we take the approximation
in Eq.~(\ref{eq:areas_v41v31}), it implies that $\vert
V_{ts}\vert$ and $\vert V_{t's}\vert$ are stronger than the
counterparts involving $d$, which implies interesting CPV effects
in $b\to s$ processes, including in $B_s \to J/\psi
\phi$~\cite{HNS05,recent,HNS07}. This is precisely where we are
finding several experimental
hints~\cite{psiphi,psiphiCDF10,ASLDzero}!
We have thus clarified the phenomenological link and reasoning
behind Eq.~(\ref{eq:J234_v41v31}), which echoes quite well those
given in Ref.~\cite{Hou09} for Eq.~(\ref{eq:J234_Hou}), but in a
more hand-waving way.

One may question whether our choice to eliminate the index ``1''
via Eq.~(\ref{eq:eliminate1}) can keep its generality, once we
introduce the hierarchy of quark masses. To check this, we note
that one could apply small mass expansion to
Eq.~(\ref{eq:Jprime_def}) directly and obtain the same results
without passing any algebra, or even using
Eq.~(\ref{eq:eliminate1}). Taking Eq.~(\ref{eq:J234_Hou}) as a
guide, by collecting terms with factor $m_{t'}^4 m_t^2$ in the
summation in $J'(2,3,4)$ in Eq.~(\ref{eq:Jprime_def}), one has
\begin{equation}
\begin{split}
J'(2,3,4) & \sim \;\,\,\, m_{t'}^4 m_t^2 m_{b'}^4 m_b^2
\left[\Vsq{t'b'}A^{tt'}_{sb}
                     -\Vsq{t'b}A^{tt'}_{sb'}
                     +\Vsq{t's}A^{tt'}_{bb'}\right], \\
J'(1,3,4) & \sim - m_{t'}^4 m_t^2 m_{b'}^4 m_b^2
\left[\Vsq{t'b'}A^{tt'}_{sb}
                     -\Vsq{t'b}A^{tt'}_{sb'}
                     +(\Vsq{t's}-1)A^{tt'}_{bb'}\right],
\end{split}
  \label{eq:J234_fast}
\end{equation}
which is exactly what we have in Eqs.~(\ref{eq:2invariants}) and
(\ref{eq:Lambda234_nomc}) when neglecting any terms of order equal
to or smaller than $m_t^2/m_{t'}^2 $. If we invoke $\vert
V_{t'd}\vert$ to be much smaller than the other three elements,
and Eq.~(\ref{eq:areas_v41v31}) is satisfied, then we again get a
formula for $J'(2,3,4)$ that is in line with
Eq.~(\ref{eq:J234_v41v31}), while $J'(1,3,4)$ cancels away as in
Eq.~(\ref{eq:J134_v41v31}).

One may then question the utility of the algebra, which
constitutes the bulk of the paper. Note that
Eq.~(\ref{eq:J234_fast}) does not satisfy the requirements for the
$t$--$t'$ degeneracy limit. This can be remedied by collecting the
$m_{t'}^2 m_t^4$ terms as well. But if one wishes to explore other
subleading effects, then Eq.~(\ref{eq:J234_fast}) offers no
guidance, and to explore these, one might as well resort to
Eqs.~(\ref{eq:4invariants}) and (\ref{eq:Lambda234_def}). As we
will show in Discussion below, it is possible, through the
structure of the CKM matrix, that Eq.~(\ref{eq:J234_v41v31})
(hence Eq.~(\ref{eq:J234_fast})) in fact is absent.
Thus, Eqs.~(\ref{eq:4invariants}) and (\ref{eq:Lambda234_def})
offer the general starting point, independent of the quark mass
hierarchy. It provided an easy way to evaluate the leading effects
with the hierarchy taken into account, and can always be used in
discussing special CKM structures. Our formulas show the complete
mass factors in front of each CKM triangle area, which all have
the form of difference of mass squares. This feature allows us to
explore some more general cases, as we will discuss in the next
section.

\section{Discussion}

At the end of the previous section, we have seen the implications
for very small $\vert V_{td}\vert$ and $\vert V_{t'd}\vert$ and
quark mass hierarchy.
Only one of the four Jarlskog invariants, $J'(2,3,4)$ of
Eq.~(\ref{eq:J234_v41v31}), remains nonzero, while the other
three are subleading, and hence could in principle vanish
if we have exact $\vert V_{td}\vert = \vert V_{t'd}\vert = 0$
and $m_u = m_c$.
In this case, it would be an effective three-generation world,
where the first generation decouples from the
other three heavier generations, when taking the extra freedom
in $\vert V_{cd}\vert$ provided by $u$--$c$ degeneracy.
This seems to be in contrast to the assertion by Jarlskog in
Ref.~\cite{Jarlskog87} that when three of these invariants vanish exactly,
the fourth would also vanish.
However, as Jarlskog mentioned, this assertion is not valid in
the present case. In fact, this assertion is not valid whenever one generation
decouples from the other three generations (zeros in the CKM matrix).

Using the small mass expansion is quite different
from taking mass degeneracy limits mathematically.
If one has mass degeneracy, extra freedom in the quark mixing matrix
must be taken into account. In Ref.~\cite{Jarlskog87},
the author also treated exact degeneracy differently to avoid
possible singularity.
Nevertheless, in our real world, we do not have any two quarks with
the same mass, so it is reasonable to consider only the smallness
of quarks but not degeneracy, though these two ways seem similar
physically.
It should be further noted that, if one applies $t$--$t'$ degeneracy,
then $J'(2,3,4)$ and $J'(1,3,4)$ do not seem to vanish, which seems paradoxical.
This can be traced, however, to Eq.~(\ref{eq:Jprime_def}), where the mass-squared
difference appears in the denominator in defining the projection operators,
and the second equality cannot apply in the $t$--$t'$ degeneracy limit.
To address this issue, rather than flipping the definition of primed versus unprimed
objects, we could inspect the behavior of $b$--$b'$ degeneracy limit instead of
$t$--$t'$. One immediately sees that, if one maintains $m_{b'} - m_b > m_s > m_d$ while
letting $m_s \to 0$, then taking $b'$--$b$ degeneracy limit, all four Jarlskog invariants
would properly vanish, hence the previous paradox is an artefact of choosing to decompose
down-type Jarlskog invariants. But we then see that ``$t$--$t'$ degeneracy"
(that mimic true $b$--$b'$ degeneracy) indeed cannot be applied.
We therefore gain an insight that, while massless degeneracy of the first two generations
can simultaneously be applied for up and down type quarks (because of vanishing mass
protection in reaching second equality of Eq.~(\ref{eq:Jprime_def})), this is
not so for the degeneracy of the massive 3rd and 4th generations.

In the previous section, we considered only the case with
small $\vert V_{td}\vert$ and $\vert V_{t'd}\vert$.
Now let us consider more scenarios when some of the elements
in the CKM matrix are extremely small,
leading to some vanishing triangle areas.
First, let us consider the case where the fourth generation
is totally decoupled from the first three generations.
One then expects an effective three-generation theory,
and the CPV effect should be the same as in SM3.
Due to the decoupling, all rotation angles which link the fourth
generation and lower generations are zero. That is,
\[ V_{14} = V_{24} = V_{34} = V_{41} = V_{42} = V_{43}=0, \]
and it follows that any $A^{u_1u_2}_{d_1d_2}$ that contains $4$ in its label
is zero. $\Lambda_{234}$ is also zero because every term in $\Lambda_{234}$
contains at least one zero factor.
The four invariants then become
\begin{equation}
J'(2,3,4) = J'(1,3,4) = J'(1,4,2) = 0, \quad
J'(1,2,3) = - v'(1,2,3) v(1,2,3)A^{23}_{23},
\end{equation}
which is exactly what we have in SM3, and
there is no CPV effect induced by fourth generation, as expected.

But if the fourth generation exists, it is hard to conceive that it
decouples from all other generations.
Consider the case where the fourth generation decouples from the first
two generations, that is,
\[ V_{14} = V_{24} = V_{41} = V_{42} =0, \]
then the only non-vanishing triangle areas in Eq.~(\ref{eq:4invariants})
are $A^{23}_{23}$ and $A^{34}_{34}$.
One can show from
Eqs.~(\ref{eq:six_relation_up}) and (\ref{eq:six_relation_down}) that
one must have either $A^{23}_{23}=A^{34}_{34}=0$ or $V_{34}=V_{43}=0$.
The first solution means there is no CPV at all,
and the third generation also decouples from the first two generation,
which contradicts experimental observation.
The second solution, on the other hand, means that the
fourth generation decouples also from the third generation,
which is the previous scenario we have just discussed.
This result shows that if the fourth generation does exist,
it must either couple with at least 2 lower generations,
or must fully decouple from all 3 lower generations.

However, even if the fourth generation is present and couples to all other
generations, it is still possible that we have only one CPV phase.
Consider, for instance, having $b'$ decoupled from $u$ and $c$, which
gives $ V_{14} = V_{24} = 0$.
Then any triangle areas $A^{u_1u_2}_{d_1d_2}$ with a ``4'' in lower
indices and an ``1'' or a ``2'' in upper indices will vanish.
In addition, there are other triangle areas that would also vanish by
using the unitarity condition,
\begin{equation}
A^{34}_{14,42,34}=-A^{14}_{14,42,34}-A^{24}_{14,42,34}=0.
\end{equation}
The only non-vanishing triangle areas used in Eq.~(\ref{eq:4invariants})
are $A^{23}_{23}$, $A^{34}_{23}$, and $A^{42}_{23}$.
But Eq.~(\ref{eq:six_relation_up}) with
$(u_{1},u_{2},u_{3},u_{4}) = (1,2,3,4)$ and $(1,3,2,4)$ gives
\begin{equation}
(\Vsq{34}-\Vsq{44})A^{34}_{23}=0, \quad
\Vsq{34}A^{23}_{23}=-\Vsq{44}A^{42}_{23}.
\end{equation}
Provided that $\Vsq{34} \neq 0$ and $\Vsq{34} \neq \Vsq{44}$,
there is only one degree of freedom in triangle areas, hence only one CPV phase.
All the Jarlskog invariants are then proportional to this area,
and the leading effect is
\begin{align}
J'(2,3,4)
&\sim - \dm{t'}{t}^2 \dm{c}{u} \dm{b'}{s} \dm{b'}{b} \dm{b}{s}
 \Vsq{tb'} A^{ct}_{sb}, \\
J'(1,3,4)
&\sim \; \,\ \dm{t'}{t}^2 \dm{c}{u} \dm{b'}{d} \dm{b'}{b} \dm{b}{d}
 \Vsq{tb'} A^{ct}_{sb},
\end{align}
which is smaller than Eq.~(\ref{eq:leadinginJ234}) due to the
factor ${m^2_{c}}$, but it is still enhanced by $\sim 10^{10}$
when compared with Eq.~(\ref{eq:JinSM3}). One sees that if the
fourth generation does not totally decouple from the other three,
it will leave its fingerprint on some CPV process(es).

For other possible scenarios, one can follow the same recipe we used.
First, the dependence of triangle areas are determined by the relations in
Eqs.~(\ref{eq:six_relation_up}) and (\ref{eq:six_relation_down}).
Then, inserting these relations into
Eqs.~(\ref{eq:4invariants}) and (\ref{eq:Lambda234_def}),
one can identify the leading effect in the corresponding scenario.
Finally, one should note that it is very unlikely to have any exact zero
in the quark mixing matrix from theoretical perspective,
and certainly not experimentally either.
These cases allow us to see the asymptotic behavior of the leading effect.

\section{Conclusion}

The formula in Eq.~(\ref{eq:J234_Hou}), as if involving just 2-3-4 generations
in a 4-generation world, would be enhanced above the 3-generation Jarlskog
invariant $J$ of Eq.~(\ref{eq:JinSM3}) by an astounding $10^{15}$ or so.
This is because the dependence on the small mass squared differences between
the two lightest generations get replaced by heavier masses at the weak scale.
The purpose of our study is to check to what extent Eq.~(\ref{eq:J234_Hou})
is the leading CPV effect in the 4-generation Standard Model.

We chose the more convenient starting point of Jarlskog's extension to 4
invariants in SM4. As we always maintain physical finite values for quark
masses and CKM mixing elements, this is equivalent in SM4 to the more
complicated Gronau, Kfir and Loewy approach. Through algebraic manipulations,
the more tedious of which are relegated to the appendices, we arrive at the
general results of Eq.~(\ref{eq:4invariants}), which depend on an algebraic
function $\Lambda_{234}$ defined in Eq.~(\ref{eq:Lambda234_def}). With full
generality, this does not offer too much insight. We then invoked the
hierarchy of physical quark masses, i.e. the aforementioned smallness of the
first two generations masses on the weak scale, to eliminate two Jarlskog
invariants, $J'(1,2,3)$ and $J'(1,4,2)$, as subleading, as well as simplify
$\Lambda_{234}$.
Invoking the empirical condition of small rotations, that
off-diagonal elements in $V$ are not larger than $\vert
V_{us}\vert$, does not simplify further the result of
Eq.~(\ref{eq:2invariants}). One needs further knowledge of
patterns of CKM elements (analogous to the mass hierarchy).
Because of recent hints in $b\to s$ processes, this cannot yet be
concluded. Instead, we found the relation of Eq. (19) between
triangle areas would hold, given that $b\to d$ transitions seem to
conform with 3 generation Standard Model. The resulting Eq. (21)
largely confirms the suggestion of Eq. (2). In fact, one could
have taken a much more efficient approach, for the purpose of
identifying the leading effect, by making small mass expansion
from the outset in Eq.~(\ref{eq:Jprime_def}), and arrive at
Eq.~(\ref{eq:J234_fast}). This retains all features of the
proposed $J_{2,3,4}^{sb}$ in Eq.~(\ref{eq:J234_Hou}), keeping to
$m_t^2/m_{t'}^2$ order, as well as order of CKM elements, which
could be as large as 0.1. We therefore see that though
$J_{2,3,4}^{sb}$ in Eq.~(\ref{eq:J234_Hou}) does seem to be the
leading term in the presence of quark mass hierarchies and small
rotation angles --- which is our world --- there should be a
myriad of subleading terms that are perhaps only 10 times smaller.

In the course of our study, we also uncovered the apparent
phenomenological condition for $J_{2,3,4}^{sb}$ in
Eq.~(\ref{eq:J234_Hou}) to be the leading term. Current data
suggest that CPV in $b\to s$ transitions, notably for
mixing-dependent CPV in $B_s \to J/\psi\phi$, could be sizable,
despite the B factory confirmation of consistency with a
3-generation source for $b\to d$ transitions (notably for
mixing-dependent CPV in $B_d \to J/\psi K_S$). Thus, $V_{td}$ and
$V_{t'd}$ seem subdued compared with $V_{ts}$ and $V_{t's}$ in
strength, respectively. In this case, we were able to derive
Eq.~(\ref{eq:J234_v41v31}) which is extremely close to
Eq.~(\ref{eq:J234_Hou}), except for very minor differences. We
thus conclude that, in general the claim of a large enhancement by
4th generation masses is true, although there would be several
terms comparable to $J_{2,3,4}^{sb}$ in Eq.~(\ref{eq:J234_Hou}).
If, however, we do discover sizable CPV effect in $B_s \to
J/\psi\phi$ that is much enhanced over SM3 expectations, then
indeed $J_{2,3,4}^{sb}$ of Eq.~(\ref{eq:J234_Hou}), or
$J^\prime(2,3,4)$ of Eq.~(\ref{eq:J234_v41v31}), is the single
leading term. But there would still be subleading terms that could
be just an order of magnitude less in strength, depending on the
strength of associated CKM elements.

\appendix

\section{Some Algebra}
\label{deriveJinA}
We start from Eq.~(\ref{eq:Jprime_def})
\begin{equation}
\label{eq:Jprime_def_app}
J'(a,b,c) = v'(a,b,c) \sum_{i,j,k=1}^4 m^2_i m^2_j m^2_k \,
            \Im(\V{i}{a}{b}\V{j}{b}{c}\V{k}{c}{a}).
\end{equation}
%
Replace every term which contains $m_1^2$ in the above summation
by the unitarity condition, the summation becomes
\begin{equation}
\label{eq:remain_J_pre}
\sum_{i,j,k=2}^4\dmu{i}{1}\dmu{j}{1}\dmu{k}{1} \,
    \Im (\V{i}{a}{b}\V{j}{b}{c}\V{k}{c}{a}),
\end{equation}
where now the sum is over all possible $i,j,k$ in the set $\{2,3,4\}$.
Note that since $a,b,c$ are all different, there exists no term like
$\delta_{aa}=1$ in Eq.~(\ref{eq:remain_J_pre}).

We can apply the similar trick to the down-type indices of $V$.
Consider the case $ a = 1 $ and $ b,c $ are chosen differently from $\{2,3,4\}$.
\begin{equation}
\label{eq:replace_down}
\begin{split}
  \Im (\V{i}{1}{b}\V{j}{b}{c}\V{k}{c}{1}) =
 - &\Im (\V{i}{2}{b}\V{j}{b}{c}\V{k}{c}{2})
 -  \Im (\V{i}{3}{b}\V{j}{b}{c}\V{k}{c}{3})\\
 - &\Im (\V{i}{4}{b}\V{j}{b}{c}\V{k}{c}{4})
 +  \Im (V_{ib}^{*}\V{j}{b}{c}V_{kc})\delta_{ik}\\
=- &\Im (\V{i}{b}{b}\V{j}{b}{c}\V{k}{c}{b})
 -  \Im (\V{i}{c}{b}\V{j}{b}{c}\V{k}{c}{c})\\
 - &\Im (\V{i}{d}{b}\V{j}{b}{c}\V{k}{c}{d})
 +  \Im (V_{ib}^{*}\V{j}{b}{c}V_{kc})\delta_{ik}\\
=- &\Im (\V{i}{d}{b}\V{j}{b}{c}\V{k}{c}{d})
 +  \Im (V_{ib}^*\V{j}{b}{c}V_{kc})\delta_{ik}\\
 + &\Vsq{ib}A^{jk}_{bc}
 -  \Vsq{kc}A^{ij}_{bc},
\end{split}
\end{equation}
where $d$ is taken to be different from $a,b,c$, and the second equality
follows from replacing $2,3,4$ by $b,c,d$, by reordering the first
three terms. Real factors are taken out in the third equality, and we also
used
\begin{equation}
\begin{split}
A^{ij}_{bc} = \Im\left[(\V{i}{b}{c})^*\V{j}{b}{c}\right].
\end{split}
\end{equation}
Substituting Eqs.~(\ref{eq:remain_J_pre}) and (\ref{eq:replace_down}) back into
Eq.~(\ref{eq:Jprime_def_app}), we have
\begin{equation}
\begin{split}
J'(1,b,c) = v'(1,b,c)\Bigg\{ &\sum_{i,j=2}^4
    \dmu{i}{1}^2\dmu{j}{1}\, A^{ij}_{bc} \\
- &\sum_{i,j,k=2}^4\dmu{i}{1}\dmu{j}{1}\dmu{k}{1} \, \Im
  (\V{i}{d}{b}\V{j}{b}{c}\V{k}{c}{d}) \\
  + & \sum_{i,j,k=2}^4\dmu{i}{1}\dmu{j}{1}\dmu{k}{1}
  ( \Vsq{ib}A^{jk}_{bc}-\Vsq{kc}A^{ij}_{bc} ) \Bigg\}.
\end{split}
\end{equation}
The third summation would vanish, since the upper indices of $A$
are anti-symmetric hence each component will cancel one another.

Define now
\begin{equation}
\label{eq:Lambda_def}
\Lambda_{dbc} = -\sum_{i,j,k=2}^4\dmu{i}{1}\dmu{j}{1}\dmu{k}{1}
 \Im(\V{i}{d}{b}\V{j}{b}{c}\V{k}{c}{d}).
\end{equation}
We note that the indices of $A$ and $\Lambda$ are antisymmetric,
and hence we have
\begin{equation}
\dmu{i}{1}^2\dmu{j}{1} A^{ij}_{bc} + \dmu{j}{1}^2\dmu{i}{1} A^{ji}_{bc}
    = - v(1,i,j) A^{ij}_{bc}.
\end{equation}
Thus the four Jarlskog invariants in SM4 can be written as
\begin{equation}
\begin{split}
J'(2,3,4) &= - v'(2,3,4) \Lambda_{234}, \\
J'(1,3,4) &=  v'(1,3,4) \left[\Lambda_{234}-v(1,3,4)A^{34}_{34}-v(1,2,3)A^{23}_{34}-v(1,4,2)A^{42}_{34} \right], \\
J'(1,2,3) &=  v'(1,2,3) \left[\Lambda_{234}-v(1,3,4)A^{34}_{23}-v(1,2,3)A^{23}_{23}-v(1,4,2)A^{42}_{23} \right], \\
J'(1,4,2) &=  v'(1,4,2) \left[\Lambda_{234}-v(1,3,4)A^{34}_{42}-v(1,2,3)A^{23}_{42}-v(1,4,2)A^{42}_{42} \right].
\end{split}
\end{equation}

$\Lambda_{234}$ can be expressed further in terms of triangle areas.
From Eq.~(\ref{eq:Lambda_def}), we consider the following quantity,
\begin{equation}
\label{eq:lambda_omega_def}
\begin{split}
- \dmu{i}{1}\dmu{j}{1}\dmu{k}{1} \Im (\V{i}{b}{c}\V{j}{c}{d}\V{k}{d}{b})=
-\mathcal{M}^{ijk}_{1}\omega^{ijk}_{bcd},
\end{split}
\end{equation}
where we have defined the mass prefactor
$\mathcal{M}^{ijk}_{1} \equiv \dmu{i}{1}\dmu{j}{1}\dmu{k}{1}$
and the imaginary part of six CKM matrix elements
$\omega^{ijk}_{bcd} \equiv\Im (\V{i}{b}{c}\V{j}{c}{d}\V{k}{d}{b})$.
Note that $\omega^{ijk}_{bcd}$ has the following properties,
\begin{equation}
\label{eq:omega_property}
\begin{split}
\sum_{i=2}^4\omega^{ijk}_{bcd}&=-\omega^{1jk}_{bcd}, \\
\omega^{ijk}_{bcd}&=\omega^{jki}_{cdb}=\omega^{kij}_{dbc}, \\
\omega^{iii}_{bcd}&=0, \\
\omega^{iji}_{bcd}&=\Vsq{ib}A^{ij}_{cd}.
\end{split}
\end{equation}
$\Lambda_{bcd}$ could be written as the following summation,
\begin{equation}
\label{eq:Lambda_sum}
\begin{split}
\Lambda_{bcd}=\sum_{i,j,k=2}^4 - \mathcal{M}^{ijk}_{1}\omega^{ijk}_{bcd}.
\end{split}
\end{equation}

Note that from the second property of Eq.~(\ref{eq:omega_property}),
$\Lambda_{bcd}=\Lambda_{dbc}=\Lambda_{cdb}$,
while the first property of Eq.~(\ref{eq:omega_property}) implies,
\begin{equation}
\label{eq:zero_sum}
\begin{split}
\sum_{i,j,k=2}^4 \omega^{ijk}_{bcd}
=\sum_{j,k=2}^4 - \omega^{1jk}_{bcd}
=\sum_{k=2}^4 \omega^{11k}_{bcd}
=-\omega^{111}_{bcd}
=0.
\end{split}
\end{equation}
Combining Eqs.~(\ref{eq:Lambda_sum}) and (\ref{eq:zero_sum}), we have
\begin{equation}
\label{eq:Lambda_bcd}
\Lambda_{bcd} = \sum_{i,j,k=2}^4
- (\mathcal{M}^{ijk}_{1}-\mathcal{M}^{234}_{1})\omega^{ijk}_{bcd}.
\end{equation}
Since the upper indices of $\mathcal{M}$ are symmetric and
$\omega^{iii}_{bcd}=0$,
$(\mathcal{M}^{ijk}_{1}-\mathcal{M}^{234}_{1})\omega^{ijk}_{bcd} = 0$
if $(i,j,k)$ are taken to be all different or all the same from $\{2,3,4\}$.
Thus, Eq.~(\ref{eq:Lambda_bcd}) could be written as
\begin{equation}
\begin{split}
\Lambda_{bcd}
&=\left[(\mathcal{M}^{234}_{1}-\mathcal{M}^{232}_{1})
    (\omega^{232}_{bcd}+\omega^{232}_{cdb}+\omega^{232}_{dbc})
       +(\mathcal{M}^{234}_{1}-\mathcal{M}^{323}_{1})
    (\omega^{323}_{bcd}+\omega^{323}_{cdb}+\omega^{323}_{dbc}) \right]\\
&\quad + \textrm{other two cyclic permutations of } (2,3,4),
\end{split}
\end{equation}
where we have used the symmetric property of the upper indices of $\mathcal{M}$
and also the second property of Eq.~(\ref{eq:omega_property}).
Using now the fourth property of Eq.~(\ref{eq:omega_property})
and express terms like $\omega^{232}_{bcd}$ in terms of triangle areas with real factors,
$\Lambda_{bcd}$ becomes
\begin{equation}
\begin{split}
\Lambda_{bcd} = &
\Big[\quad \ \dmu{2}{1}\dmu{3}{1}\dmu{4}{2}
    (\Vsq{2b}A^{23}_{cd}+\Vsq{2c}A^{23}_{db}+\Vsq{2d}A^{23}_{bc})\\
& \; \ - \dmu{2}{1}\dmu{3}{1}\dmu{4}{3}
    (\Vsq{3b}A^{23}_{cd}+\Vsq{3c}A^{23}_{db}+\Vsq{3d}A^{23}_{bc})\Big]\\
&+ \textrm{other two cyclic permutations of } (2,3,4).
\end{split}
\end{equation}
Substituting $(2,3,4)$ for $(b,c,d)$
one obtains Eq.~(\ref{eq:Lambda234_def}).

\section{Relations between Triangle Areas}
\label{six_area_relation}

Since there are only three independent phases in the CKM matrix in SM4,
one expects there exists some relations among the nine triangle areas.
For example, one can obtain $J'(1,3,4)$ directly from $J'(2,3,4)$ by exchanging
the indices 1 and 2, rather than using the approach we presented.
In this case, one would get two different expressions for $J'(1,3,4)$.
By comparing the two, one would find non-trivial relations of some
of the triangle areas.

There are in fact 6 relations, consisting of three up-type relations,
\begin{equation}
\label{eq:six_relation_up}
\left[(\Vsq{u_{1}n}-\Vsq{u_{2}n})A^{u_{1}u_{2}}_{lm}
+(\Vsq{u_{3}n}-\Vsq{u_{4}n})A^{u_{3}u_{4}}_{lm} \right]
+ \textrm{cyclic permu.~of } (l,m,n)  =0,
\end{equation}
and three down-type relations,
\begin{equation}
\label{eq:six_relation_down}
\left[(\Vsq{nd_{1}}-\Vsq{nd_{2}})A^{lm}_{d_{1}d_{2}}
+(\Vsq{nd_{3}}-\Vsq{nd_{4}})A^{lm}_{d_{3}d_{4}}\right]
+ \textrm{cyclic permu.~of } (l,m,n) =0,
\end{equation}
where $u_{1}$ to $u_{4}$ (or $d_{1}$ to $d_{4}$ for down-type)
are taken all differently from $\{1,2,3,4\}$,
and $l, m, n$ are also taken all differently from $\{1,2,3,4\}$
Different choices of $(l,m,n)$ would in fact give equivalent relations,
so throughout this appendix we will regard $(l,m,n)$ as given labels,
say $(2,3,4)$, without loss of generality.

Let us present a direct proof of these 6 relations.
First, we consider the up-type relations.
We define the LHS of Eq.~(\ref{eq:six_relation_up}) to be
\begin{equation}
\label{eq:six_relation_absorb}
\begin{split}
\mathcal{R}_{u_1u_2u_3u_4}=
&[-\Im(\V{u_{1}}{n}{n}\V{u_{1}}{l}{m}\V{u_{2}}{m}{l})
  -\Im(\V{u_{2}}{n}{n}\V{u_{2}}{l}{m}\V{u_{1}}{m}{l}) \\
 &-\Im(\V{u_{3}}{n}{n}\V{u_{3}}{l}{m}\V{u_{4}}{m}{l})
  -\Im(\V{u_{4}}{n}{n}\V{u_{4}}{l}{m}\V{u_{3}}{m}{l})]\\
 &+\textrm{other two cyclic permutations of } (l,m,n),
\end{split}
\end{equation}
where we put back the real factors into $\Im(\cdots)$.
With the unitarity condition, we can replace
$\V{u_{1}}{n}{l}$ by $-\V{u_{2}}{n}{l}-\V{u_{3}}{n}{l}-\V{u_{4}}{n}{l}$,
and substitute this into Eq.~(\ref{eq:six_relation_absorb}).
For example, the first term in Eq.~(\ref{eq:six_relation_absorb}) would become
\begin{equation}
\begin{split}
-\Im(\V{u_{1}}{n}{n}\V{u_{1}}{l}{m}\V{u_{2}}{m}{l})
=&\Vsq{u_{2}l}\Im(\V{u_{1}}{n}{m}\V{u_{2}}{m}{n}) \\
 &+\Im(\V{u_{3}}{l}{n}\V{u_{1}}{n}{m}\V{u_{2}}{m}{l}) \\
 &+\Im(\V{u_{4}}{l}{n}\V{u_{1}}{n}{m}\V{u_{2}}{m}{l})\\
=&\Vsq{u_{2}l}A^{u_{1}u_{2}}_{mn}
  +\omega^{u_{3}u_{1}u_{2}}_{lnm}+\omega^{u_{4}u_{1}u_{2}}_{lnm},
\end{split}
\end{equation}
where $\omega^{u_{3}u{1}u_{2}}_{lnm}$ is defined in Eq.~(\ref{eq:lambda_omega_def}).
Applying the unitarity condition to every term in Eq.~(\ref{eq:six_relation_absorb}),
we have
\begin{equation}
\begin{split}
\mathcal{R}_{u_1u_2u_3u_4} &= \Big[
    \Vsq{u_{2}l}A^{u_{1}u_{2}}_{mn}-\Vsq{u_{1}l}A^{u_{1}u_{2}}_{mn}
    +\Vsq{u_{3}l}A^{u_{3}u_{3}}_{mn}-\Vsq{u_{4}l}A^{u_{3}u_{4}}_{mn} \\
&\quad  +\omega^{u_{3}u_{1}u_{2}}_{lnm}+\omega^{u_{4}u_{1}u_{2}}_{lnm}
    +\omega^{u_{3}u_{2}u_{1}}_{lnm}+\omega^{u_{4}u_{2}u_{1}}_{lnm}
    +\omega^{u_{1}u_{3}u_{4}}_{lnm}+\omega^{u_{2}u_{3}u_{4}}_{lnm}
    +\omega^{u_{1}u_{4}u_{3}}_{lnm}+\omega^{u_{2}u_{4}u_{3}}_{lnm} \Big]\\
&\quad  +\textrm{other two cyclic permutations of } (l,m,n).
\end{split}
\end{equation}
The first line in the brackets in the RHS together with the cyclic permutations
amounts to $-\mathcal{R}_{u_1u_2u_3u_4} $.
And due to the second property of Eq.~(\ref{eq:omega_property}),
the cyclic permutations on $\omega$'s lower labels
can be moved to its upper labels when all cyclic permutations are summed,
so the left 24 $\omega$'s can be written as
$ \sum_{(i,j,k)'}\omega^{ijk}_{lnm} $
where $(i,j,k)'$ are taken all differently from $\{u_{1},u_{2},u_{3},u_{4}\}$,
which is equivalent to the set $\{1,2,3,4\}$.
So now we have
\begin{equation}
\label{eq:R_subtraction_extra}
\begin{split}
2 \mathcal{R}_{u_1u_2u_3u_4} &= \sum_{(i,j,k)'} \omega^{ijk}_{lnm}\\
&= \sum_{i,j,k = 1}^4 \omega^{ijk}_{lnm}
 -\sum_{i,j = 1}^4
    (\omega^{iij}_{lnm} + \omega^{iji}_{lnm} + \omega^{jii}_{lnm}).
\end{split}
\end{equation}
In the second equality, we allow all possible $i,j,k$ in the summation
and subtract back the extra terms.
However, from the first property of Eq.~(\ref{eq:omega_property}),
$ \sum_{a=1}^4\omega^{abc}_{lnm}=0 $,
so all terms in the RHS of Eq.~(\ref{eq:R_subtraction_extra}) vanish
when summing over $j$ from 1 to 4. Finally we have
\begin{equation}
\begin{split}
\mathcal{R}_{u_1u_2u_3u_4}  = 0,
\end{split}
\end{equation}
which proves Eq.~(\ref{eq:six_relation_up}).
The down-type relations Eq.~(\ref{eq:six_relation_down}) can be derived similarly.

As mentioned in \cite{Botella86},
in SM4 there are nine independent CKM triangle areas,
but the relations shown in this appendix seem to reduce the number of
independent CKM triangle areas to at most three.
There is in fact no conflict.
Provided the magnitude of each CKM mixing matrix element is known,
the degree of freedom of the CKM triangles
can be further reduced to at most three,
which equals the number of physical phases,
as shown in this appendix.
This result satisfies our intuition because
one needs rotation angles and phases to describe the mixing matrix.
So if one does not know the magnitude of matrix elements,
the degree of freedom of triangle areas would be larger than
the number of physical phases, but when those rotation angles are known,
one should be able to express some CKM triangle areas in terms of others
like what we proposed in Eqs.~(\ref{eq:six_relation_up}) and (\ref{eq:six_relation_down}).

\end{document}